\documentclass[epj,referee,amsmath,amssymb]{svjour}
\usepackage{latexsym}
\usepackage{graphics}
\begin{document}
\title{Shubnikov-de Haas oscillations spectrum of the strongly correlated quasi-2D organic metal (ET)$_8$[Hg$_4$Cl$_{12}$(C$_6$H$_5$Br)]$_2$ under pressure.}
\author{David Vignolles\inst{1} \and Alain Audouard\inst{1} \and Rustem B. Lyubovskii\inst{2} \and Marc Nardone\inst{1}
\and Enric Canadell\inst{3} \and Elena I. Zhilyaeva\inst{2} \and Rimma N. Lyubovskaya\inst{2}% etc
}                     % Do not remove
\mail{audouard@lncmp.org}
\institute{Laboratoire National des Champs Magn\'{e}tiques
Puls\'{e}s \thanks{UMR 5147: Unit\'{e} Mixte de Recherche CNRS -
Universit\'{e} Paul Sabatier - INSA de Toulouse}, 143 avenue de
Rangueil, 31400 Toulouse, France.  \and Institute of Problems of
Chemical Physics, RAS, 142432 Chernogolovka, MD, Russia. \and
Institut de Ci\`{e}ncia de Materials de Barcelona, CSIC, Campus de
la UAB, 08193, Bellaterra, Spain.}
\date{Received: \today / Revised version: date}
\abstract{Pressure dependence of the Shubnikov-de Haas (SdH)
oscillations spectra of the quasi-two dimensional organic metal
(ET)$_8$[Hg$_4$Cl$_{12}$(C$_6$H$_5$Br)]$_2$ have been studied up
to 1.1 GPa in pulsed magnetic fields of up to 54 T. According to
band structure calculations, its Fermi surface can be regarded as
a network of compensated orbits. The SdH spectra exhibit many
Fourier components typical of such a network, most of them being
forbidden in the framework of the semiclassical model. Their
amplitude remains large in all the pressure range studied which
likely rules out chemical potential oscillation as a dominant
contribution to their origin, in agreement with recent
calculations relevant to compensated Fermi liquids. In addition to
a strong decrease of the magnetic breakdown field and effective
masses, the latter being likely due to a reduction of the strength
of electron correlations, a sizeable increase of the scattering
rate is observed as the applied pressure increases. This latter
point, which is at variance with data of most charge transfer
salts is discussed in connection with pressure-induced features of
the temperature dependence of the zero-field interlayer
resistance.
  % end of PACS codes
} %end of abstract
\authorrunning{D. Vignolles et al.}
\titlerunning{SdH oscillations in (ET)$_8$[Hg$_4$Cl$_{12}$(C$_6$H$_5$Br)]$_2$ under pressure}
\maketitle
\section{Introduction}
\label{intro}

Quantum oscillations spectra of quasi-two-dimensional (q-2D)
multiband metals are known to contain many Fourier components. In
addition to those that can be attributed to closed orbits,
eventually induced by magnetic breakdown (MB), linear combinations
of few basic frequencies that are not predicted by the
semiclassical model \cite{Fa66,Sh84} can nevertheless be observed.
This frequency mixing phenomenon can be linked to both the q-2D
nature of the dispersion relation which is liable to give rise to
a field-dependent oscillation of the chemical potential \cite{mu}
and to the Fermi surface (FS) topology that can be regarded as a
network of orbits coupled by MB \cite{LLB}. Despite both
theoretical and experimental efforts, the latter being mainly
focused on organic metals based on the ET molecule (where ET
stands for bisethylenedithia-tetrathiofulvalene) with the FS
originating from the hybridization of one orbit with an area equal
to that of the first Brillouin zone (FBZ), quantitative
interpretation of the data is still an open problem.

Networks of orbits can also result from the hybridization of two
or more pairs of q-1D sheets as it is the case of the room
temperature FS of numerous oxide bronzes. However hidden nesting
properties \cite{Wh89} lead to the condensation of a charge
density wave which strongly modifies the FS at low temperature in
most cases. As a result, quantum oscillations spectra observed e.
g. in monophosphate tungsten bronzes \cite{Be00} can hardly be
reconciled with band structure calculations \cite{Fo02}. Analogous
problematic can be observed in the q-2D organic metal
$\beta$''-(ET)(TCNQ) \cite{Ya}.

On the contrary, although the FS of the q-2D charge transfer salt
(ET)$_8$Hg$_4$Cl$_{12}$(C$_6$H$_5$Cl), noted hereafter as (Cl,
Cl), originates from the hybridization of two pairs of q-1D sheets
as well, it remains metallic down to the lowest temperatures.
According to band structure calculations \cite{Ve94}, its FS is
composed of two compensated orbits labelled $a$ in the following
with an area of 13 \% of the FBZ area. Shubnikov-de Haas (SdH)
oscillations spectra of this compound \cite{Pr02}, which otherwise
exhibit many frequency combinations, as discussed below, are in
good agreement with this picture since the main Fourier component
has a frequency F$_a$ = 241.5 $\pm$ 2 T that corresponds to 11 \%
of the FBZ area. This is also the case of the isostructural
compound (ET)$_8$[Hg$_4$Cl$_{12}$(C$_6$H$_5$Br)]$_2$, noted
hereafter as (Cl, Br), which has been even more extensively
studied at high magnetic field \cite{Vi03,Au05} although its FS
has not been reported up to now.

With respect to the interpretation of frequency combinations,
applied pressure can be useful in order to tune both the
transverse interactions (which are involved in the chemical
potential oscillations) \cite{Ki02} and the MB gaps. In a first
step, band structure calculations of the (Cl, Br) compound at
ambient pressure are reported which confirm that its FS can be
regarded as a network of compensated orbits. In a second step, the
temperature dependence of the interlayer resistance under pressure
is considered. Finally, the pressure-induced change of the SdH
oscillations spectrum, MB field, effective masses and scattering
rate is investigated up to 1.1 GPa.

\section{Experimental}
The studied crystals were prepared by the electrochemical method
reported in \cite{Ly91}. Tight binding band structure
calculations, based on X-ray data collected at room temperature
and ambient pressure, were performed using the same method as
reported in \cite{Ve94}. The calculations use an extended
H\"{u}ckel effective one-electron Hamiltonian \cite{Wh78}. The
off-diagonal matrix elements of the Hamiltonian were calculated
according to the modified Wolfsberg-Helmholz formula \cite{Am78}.
All valence electrons were explicitly taken into account in the
calculations and the basis set consisted of double-$\zeta$
Slater-type orbitals for all atoms, except hydrogen.

Interlayer zero-field resistance and magnetoresistance
measurements were performed on crystals with approximate
dimensions 0.5$\times$0.5$\times$0.1 mm$^3$. Electrical contacts
to the crystal were made using annealed platinum wires of 20
$\mu$m in diameter glued with graphite paste. Alternating current
(1 $\mu$A, 77 Hz) and (5 to 10 $\mu$A, 20 kHz) was injected
parallel to the \emph{a}* direction for measurements of the
interlayer zero-field resistance and magnetoresistance,
respectively. Hydrostatic pressure was applied up to 1.1 GPa in an
anvil cell designed for isothermal measurements in pulsed magnetic
fields \cite{Na01}. In the following, the pressure applied at room
temperature is considered although a slight crystal size-dependent
pressure loss on cooling cannot be excluded \cite{Vi06}.
Magnetoresistance experiments were performed up to 54 T in pulsed
magnetic field with pulse decay duration of 0.36 s, in the
temperature range from 1.5 K to 4.2 K. Magnetic field was applied
normal to the conducting (\emph{bc}) plane. A lock-in amplifier
with a time constant of 30 $\mu$s was used to detect the signal
across the potential contacts. Analysis of the oscillatory
magnetoresistance is based on discrete Fourier transforms of the
data, calculated with a Blackman window.

\begin{figure}                                                          % Fig. 1
\centering \resizebox{\columnwidth}{!}{\includegraphics*{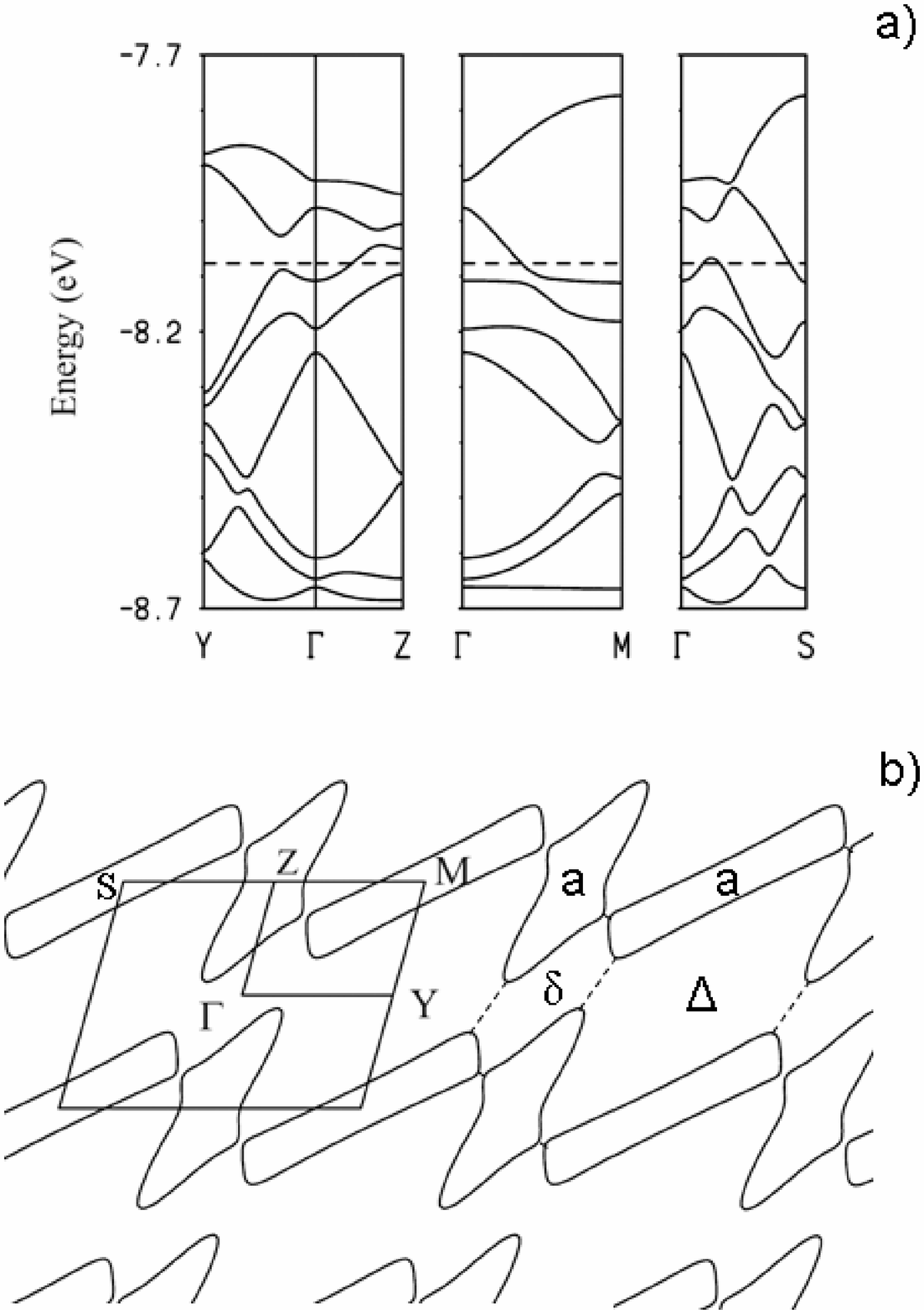}}
\caption{(a) Dispersion relations at ambient pressure of
(ET)$_8$[Hg$_4$Cl$_{12}$(C$_6$H$_5$Br)]$_2$. The dashed lines
marks the Fermi level. (b) Fermi surface (FS) corresponding to
(a). $\Gamma$, Y, Z, S and M refer to (0, 0), (b$^*$/2, 0), (0,
c$^*$/2), (-b$^*$/2, c$^*$/2) and (b$^*$/2, c$^*$/2),
respectively. Labels \emph{a}, $\delta$ and $\Delta$ correspond to
the FS pieces discussed in the text.} \label{fig_SF}
\end{figure}

\section{Results and discussion}
\label{sec_results}

The band structure calculations and FS of the (Cl, Br) compound
displayed in Fig. \ref{fig_SF} are based on crystallographic data
recorded at ambient pressure and room temperature. The repeat unit
of the donor layer contains 8 ET molecules so that the band
structure of Fig. \ref{fig_SF}a contains 8 bands mainly based on
the highest occupied molecular orbitals of ET. With the usual
oxidation states of Cl$^-$ and Hg$^{2+}$, the average charge of
the ET donors is +1/2. Consequently, two bands in Fig.
\ref{fig_SF}a should be formally empty. Since the second and third
bands from the top overlap, the system must be metallic and its FS
must contain both electron and hole contributions with the same
area. More precisely, as it is the case for (Cl, Cl) \cite{Ve94},
the FS (see Fig. \ref{fig_SF}b) originates from two pairs of
crossing q-1D sheets and constitutes a network of compensated
electron (around M) and hole (around Z) closed orbits elongated
along the $(b^* + c^*)$ and $c^*$ directions, respectively. These
orbits are both labelled \emph{a} hereafter since they have the
same area. Both the band structure and FS of Fig. \ref{fig_SF} are
very similar to those previously reported for the (Cl, Cl)
compound \cite{Ve94}. For instance whereas the area of the closed
orbits of the (Cl, Br) compound is 16.1 \% of the FBZ area,
according to data in Fig. \ref{fig_SF}, calculations for (Cl, Cl)
using the same computational details as in the present case lead
to an area of 16.7 \% of the FBZ area. Although this is not
surprising since the two compounds are isostructural, these
results justify a posteriori the previous data analysis based on a
FS topology analogous to that of the (Cl, Cl) compound
\cite{Vi03,Au05,Ly95}. Let us note that similar calculations led
to a more substantial variation of the orbits area for the (Br,
Cl) compound (9.6 \% of the FBZ area). This suggests that
replacement of Br by Cl in the anion layer has a considerably
stronger effect on the FS than the same replacement in the solvent
molecules. The small difference in orbits area between the (Cl,
Br) and (Cl, Cl) compounds is in agreement with the experimental
results. Nevertheless, a slight discrepancy between calculations
and quantum oscillations spectra can be observed since the closed
orbits area of (Cl, Br) amounts to 16.1 $\%$ and 11 $\%$ of the
FBZ area in the former and latter cases, respectively.

Zero-field interlayer resistance and magnetotransport were studied
on three crystals. Since they yield consistent results, we
concentrate on the most extensively studied one in the following.
It should be noted that all the features reported hereafter are
reversible as the pressure is released.

\begin{figure}                                                          % Fig.2
\centering \resizebox{\columnwidth}{!}{\includegraphics*{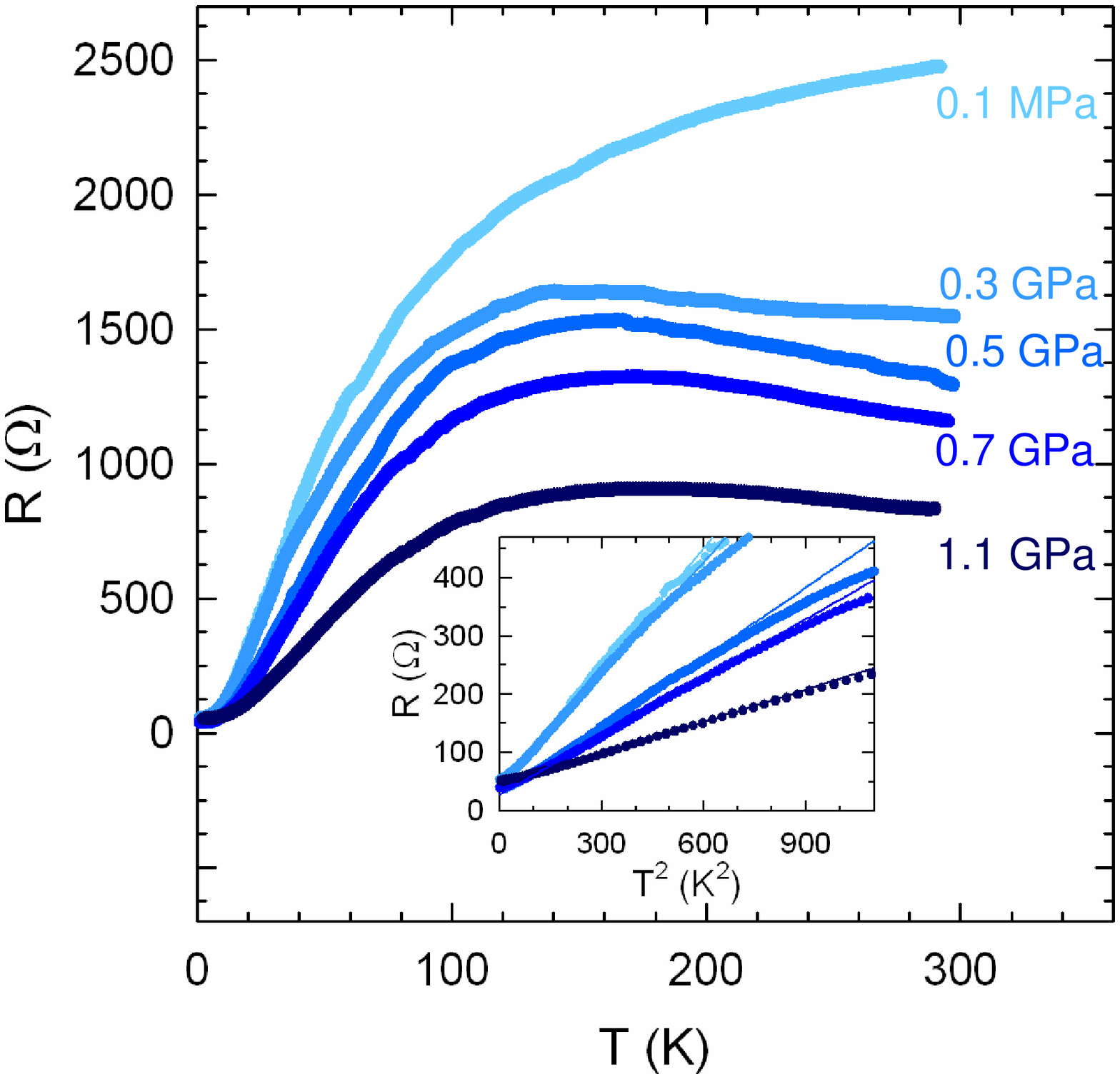}}
\caption{(color on line) Temperature dependence of the zero-field
interlayer resistance at various applied pressures. The low
temperature part of the data is plotted as a function of T$^2$ in
the inset.} \label{fig_RT}
\end{figure}

Temperature dependence of the zero-field interlayer resistance at
various applied pressures is displayed in Fig. \ref{fig_RT}. The
room temperature resistance decreases as the pressure increases
(dlnR/dP = -1 GPa$^{-1}$) while the low temperature value remains
almost unaffected. As a result, the residual resistivity ratio
(RRR) decreases monotonously by about a factor of 3 as the
pressure increases up to 1.1 GPa. Whereas the resistance
continuously decreases as the temperature decreases at ambient
pressure, a resistance maximum is observed under pressure, the
amplitude of which is maximum around 0.5 GPa. Data under pressure
share similarities with the sample-dependent zero-field interlayer
resistance of the high-resistance variant of
$\kappa$-(ET)$_2$Cu[N(CN)$_2$]Br whose resistance increases as the
temperature decreases down to about 100 K and strongly decreases
at lower temperatures \cite{St05}. A similar temperature
dependence is also observed for other charge transfer salts such
as $\kappa$-(ET)$_2$Cu(NCS)$_2$ \cite{UrMu} or
$\kappa$-(ET)$_2$Cu[N(CN)$_2$]Cl \cite{Li03}. As reported in Ref.
\cite{St99}, the resistance maximum observed in
$\kappa$-(ET)$_2$Cu[N(CN)$_2$]Br is certainly related to disorder
and possibly to point defects. Indeed, due to a decrease of the
Dingle temperature, the SdH oscillations amplitude increases as
the amplitude of the resistance maximum decreases. It should be
noted that the effect of pressure on the zero-field transport
properties of this compound is at variance with that of Fig.
\ref{fig_RT} since the amplitude of the resistance maximum
decreases continuously as the applied pressure increases in the
former case. In other words, data in zero-field could suggest
that, even though the scattering rate in
$\kappa$-(ET)$_2$Cu[N(CN)$_2$]Br decreases under pressure, it
increases in the case of the (Cl, Br) compound, at least up to 0.5
GPa. Oppositely, X-ray irradiation of $\kappa$-(ET)$_2$Cu(NCS)$_2$
lead to a decrease of the resistance maximum amplitude
\cite{An06}. However, as far as $\kappa$-(ET)$_2$Cu[N(CN)$_2$]Cl
is concerned, this behaviour is due to a significant
irradiation-induced increase of the carrier concentration
\cite{Sa08} that cannot hold in our case. It should be also
mentioned that a coherent-incoherent crossover of small polarons,
driven by the strength of the electron-phonon coupling, can also
lead to interlayer resistance maximum \cite{Lu03,Ho05}.
Alternatively, electron correlations have been invoked in order to
account for such resistance behaviour \cite{Li03,Me00}. As a
matter of fact, a T$^2$ variation of the resistance (R = R$_0$ +
AT$^2$) is observed at low temperature (see the inset of Fig.
\ref{fig_RT}). Owing to the crystal dimensions, A ranges from
$\sim$ 0.2 $\Omega$cmK$^{-2}$ at ambient pressure to $\sim$ 0.05
$\Omega$cmK$^{-2}$ at 1.1 GPa. Such a behaviour which is commonly
observed both in charge transfer salts \cite{St05,Li03} and
inorganic low dimensional compounds such as Sr$_2$RuO$_4$
\cite{Ma97} is currently regarded as the signature of a strongly
correlated Fermi liquid. This latter point and the above statement
regarding the pressure dependence of the scattering rate are
further discussed at the light of the magnetoresistance data
reported hereafter.

\begin{figure}                                                         % Fig.3
\centering
\resizebox{\columnwidth}{!}{\includegraphics*{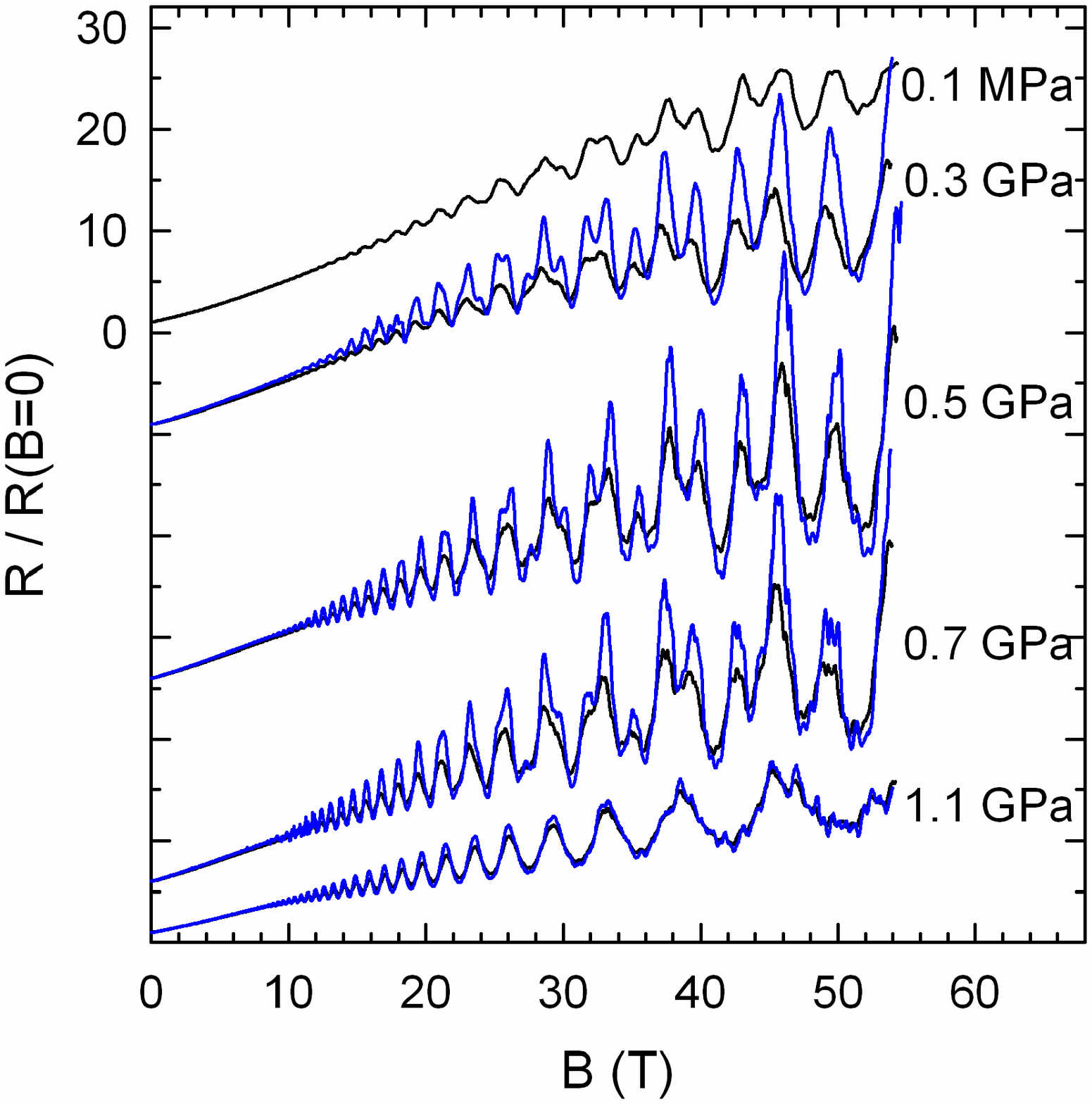}}
\caption{(color on line) Field-dependent interlayer resistance at
various applied pressures. Curves have been shifted down by an
arbitrary amount for clarity. Black and blue solid lines
correspond to data measured at 4.2 K and 1.7 K, respectively.}
\label{fig_R(B)}
\end{figure}

\begin{figure}                                                         % Fig.4
\centering \resizebox{\columnwidth}{!}{\includegraphics*{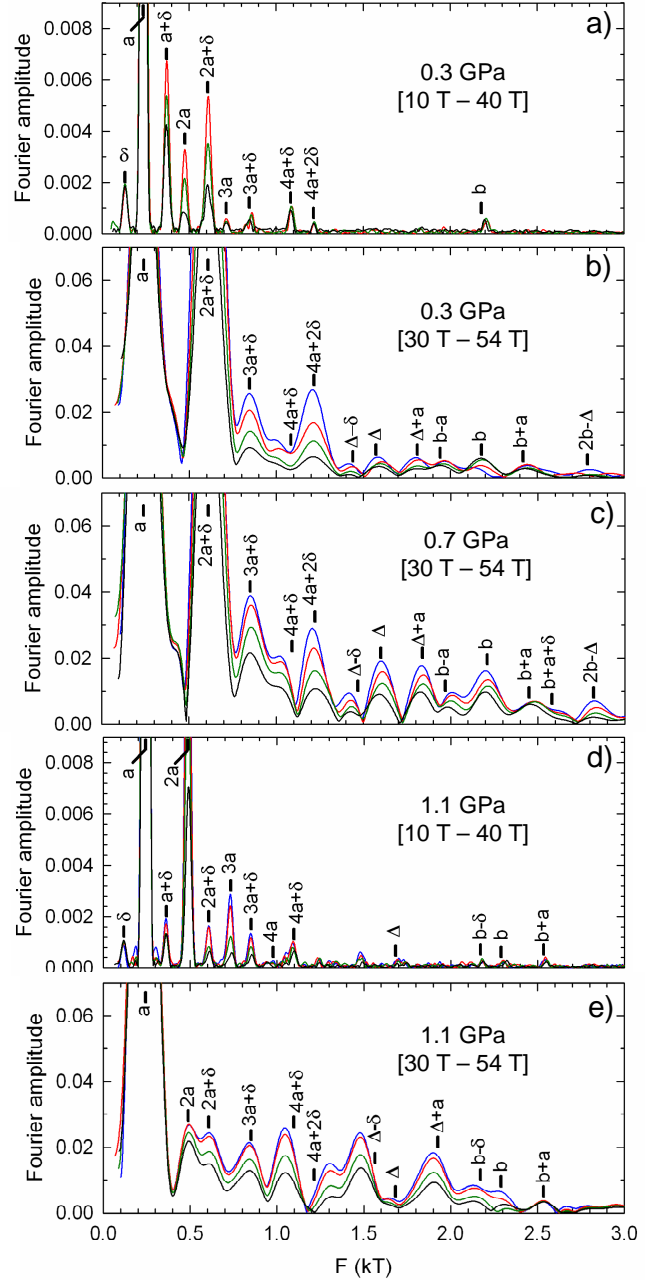}}
\caption{(color on line) Fourier analysis of the oscillatory part
of the magnetoresistance at various applied pressures and field
ranges. Blue, red, green and black solid lines correspond to data
at 1.7, 2.5, 3.4 and 4.2 K, respectively. Black vertical lines are
marks calculated with the set of frequency values (F$_a$,
F$_\delta$ and F$_\Delta$) that best fits to the Fourier spectra.}
\label{fig_TF}
\end{figure}

\begin{figure}                                                         % Fig.5
\centering
\resizebox{\columnwidth}{!}{\includegraphics*{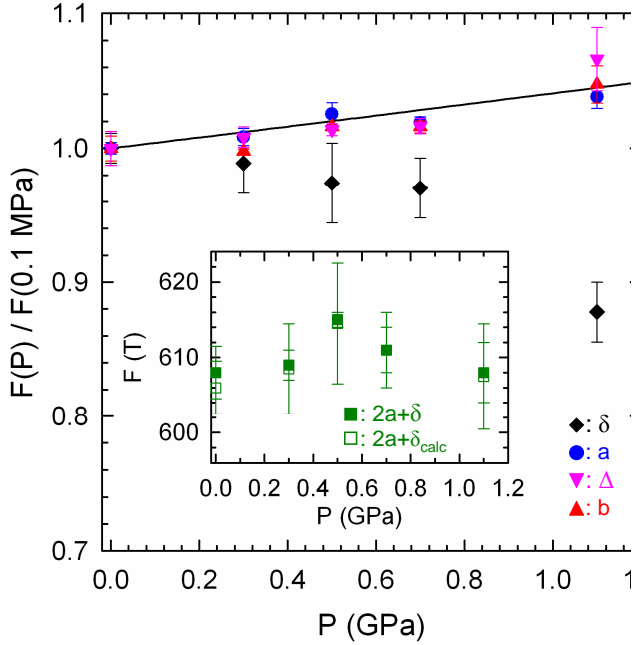}}
\caption{(color on line) Pressure dependence of  the normalized
frequency of the Fourier components $\delta$, $a$, $\Delta$ and
$b$ deduced from data in Fig. \ref{fig_TF}). The ambient pressure
values are F$_{\delta}$ = 135 T, F$_a$ = 235.5 T and  F$_{\Delta}$
= 1577 T \cite{Vi03}. The inset displays the pressure dependence
of F$_{2a + \delta}$. Open squares stand for values calculated as
F$_{2a + \delta}$ = 2F$_{a}$ + F$_{\delta}$.} \label{fig_F(P)}
\end{figure}

\begin{figure}                                                         % Fig.6
\centering
\resizebox{\columnwidth}{!}{\includegraphics*{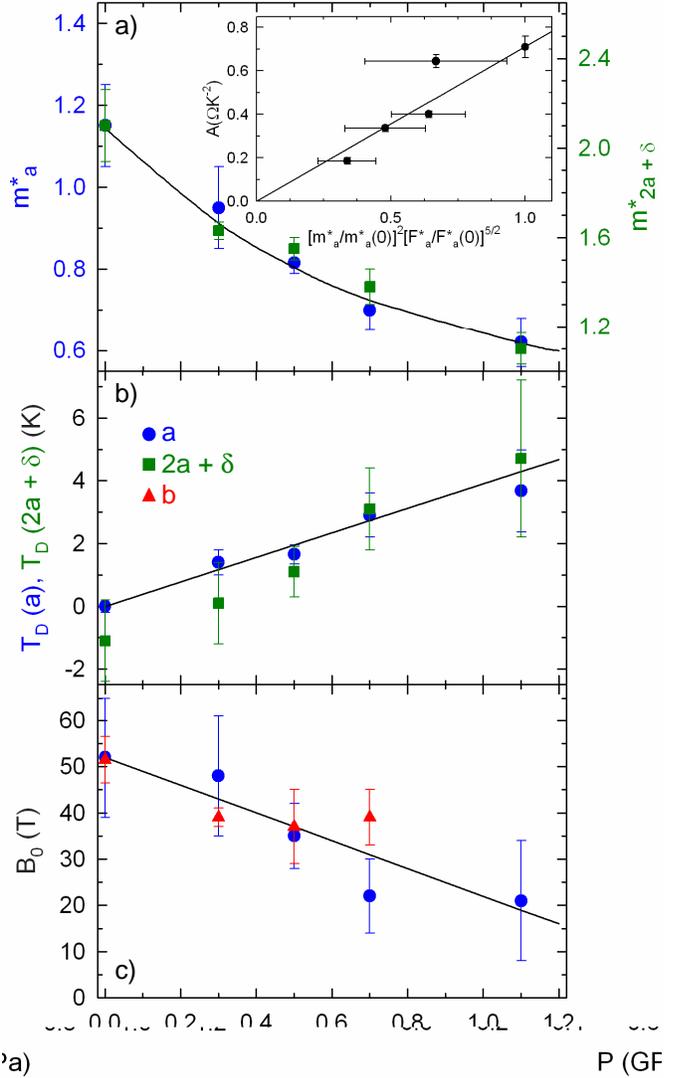}}
\caption{Pressure dependence of  the (a) effective mass, (b)
Dingle temperature and (c) magnetic breakdown field, deduced from
the field or temperature dependence of the Fourier components $a$,
$2a + \delta$ and $b$. Lines are guides to the eye. The inset
displays the A coefficient deduced from data in Fig. \ref{fig_RT},
plotted against [m$^*_a$ / m$^*_0$]$^2$ / [F$_a$ /
F$^*_0$]$^{5/2}$, where m$^*_0$ and F$^*_0$ are the ambient
pressure values. The straight line is the best fit of Eq.
\ref{Eq_A} to the data.} \label{fig_mc_TD_B0(P)}
\end{figure}

Fourier analysis of interlayer magnetoresistance data (few
examples of which are given in Fig. \ref{fig_R(B)}) are displayed
in Fig. \ref{fig_TF}. As previously reported for ambient pressure
data \cite{Vi03}, many frequencies are observed in all the
explored pressure range. Most of them are linear combinations of
that linked to the compensated electron- and hole-type orbits
(labelled $a$ in Fig. \ref{fig_SF}) and of the FS pieces located
in-between (labelled $\delta$ and $\Delta$ in Fig. \ref{fig_SF}).
These two latter FS pieces correspond to so called "forbidden
orbits"\footnote{As reported in Ref. \cite{Vi03}, these FS pieces
could correspond to MB-induced closed orbits or QI paths. However,
the corresponding effective mass and MB damping factor would be
much too large and small, respectively, to account for the data}
as it is the case of most of the observed frequency combinations.
Oppositely, the Fourier component $2a + \delta$ corresponds to a
MB orbit while the $b$ frequency corresponds to QI paths involving
an area equal to that of the FBZ (F$_b$ = 2F$_a$ + F$_{\delta}$ +
F$_{\Delta}$) \cite{Pr02,Vi03}. At high field and (or) high
pressure, shift of few frequency combinations (such as 4$a$ +
$\delta$ or $\Delta$ - $\delta$) that could as well correspond to
additional frequencies can be observed in Figs. \ref{fig_TF}(b) -
(e). Although more data is needed in order to discuss the origin
of these latter features, it can be inferred that they could arise
from slight change of the FS. Nevertheless, most of the observed
Fourier components remain linear combinations of the three basic
frequencies reported above which allows for the study of their
pressure dependence. As displayed in Fig. \ref{fig_F(P)}, F$_{b}$
(i. e. the FBZ area) linearly increases in the explored pressure
range. The pressure sensitivity (d[F$_{b}$/F$_{b}$(P =
0.1~MPa)]/dP = 0.04 GPa$^{-1}$) is in good agreement with data of
the FBZ area of numerous ET salts \cite{Ca94,Br95,Ka95}. While the
pressure dependence of F$_{a}$ is the same as that of the FBZ
area, F$_{\delta}$ strongly decreases as the pressure increases.
This behaviour, which lead to the non-monotonous pressure
dependence of F$_{2a+\delta}$ reported in the inset of Fig.
\ref{fig_F(P)}, could be understood assuming that the long and
small axis of the \emph{a} orbits elongates and shrinks,
respectively, as the pressure increases. However, although this is
actually the case of the $\beta$ orbits of $\kappa$-salts
\cite{Ca94}, such a behavior should lead to an increase of the
$\Delta$ piece area steeper than observed in Fig. \ref{fig_F(P)}.
The actual scenario is therefore certainly more complex.

The MB field (B$_0$), effective mass (m$^*$) and Dingle
temperature (T$_D$ = $\hbar$/2$\pi$k$_B\tau_D$, where $\tau_D$ is
the relaxation time) can be extracted from the field and (or)
temperature dependence of the oscillations amplitude. In the
framework of the Lifshits-Kosevich model \cite{Sh84}, the
amplitude of the Fourier component with the frequency F$_i$ is
given by A$_{i} \propto$ R$_{Ti}$R$_{Di}$R$_{MBi}$R$_{Si}$, where
the spin damping factor (R$_{Si}$) depends only on the direction
of the magnetic field with respect to the conducting plane. The
thermal (for a 2D FS), Dingle and MB damping factors are
respectively given by:

\begin{eqnarray}
\label{Rt}R_{Ti} =
\frac{{\alpha}T{}m{_i^*}}{Bsinh[{\alpha}T{}m{_i^*}/B ]}
\\R_{Di} =
exp[-{\alpha}T_{Di}m{_i^*}/B]\\
\label{RMB} R_{MBi} =
exp(-\frac{t_iB_{MB}}{2B})[1-exp(-\frac{B_{MB}}{B})]^{b_i/2}
\end{eqnarray}

where $\alpha$ = 2$\pi^2$m$_e$k$_B$/e$\hbar$ ($\simeq$ 14.69 T/K).
Integers $t_i$ and $b_i$ are respectively the number of tunnelling
and Bragg reflections encountered along the path of the
quasiparticle.

As discussed in Refs. \cite{Vi03,Au05}, only Fourier components
$a$, $2a + \delta$ (SdH) and $b$ (QI) can be analyzed on the basis
of closed orbits or QI paths, the other being due to, or strongly
affected by, the frequency mixing phenomenon. Therefore, we will
mainly focus on these oscillations in the following.

In the field range below $\sim$ 30 T, the amplitude of the Fourier
component linked to $F_b$ remains temperature-independent up to
0.7 GPa (at 1.1 GPa, the signal-to-noise ratio is too small to
derive a reliable temperature dependence). This feature is in
agreement with the zero-effective mass value  predicted for a
symmetric quantum interferometer \cite{St71}, as reported for
ambient pressure data \cite{Vi03}. Oppositely, m$^*$($a$) and
m$^*$($2a+\delta$) decreases by roughly a factor of 2 between
ambient pressure and 1.1 GPa (see Fig. \ref{fig_mc_TD_B0(P)}a).
Remarkably, the ratio m$^*$($2a+\delta$) / m$^*$($a$) remains
pressure-independent and equal to 1.8. This value is close to 2 as
expected from the semiclassical model of Falicov-Stachowiak
\cite{Fa66} for the considered coupled orbits networks
\cite{Vi03}. The decrease of the effective mass as the applied
pressure increases can be considered in the light of the T$^2$
dependence of the zero-field resistance observed at
low-temperature (see the inset of Fig. \ref{fig_RT}). Indeed, in
the case of a strongly correlated Fermi liquid, the Kadowaki-Woods
ratio (A / $\gamma^2$, where A and $\gamma$ are the T$^2$
coefficient of the resistivity and the electronic specific heat
coefficient, respectively) should be proportional to the lattice
parameter \cite{Mo95}. Assuming, as a rough approximation, that
$\gamma \propto 1 / \varepsilon_F \propto$ m$^*$($a$) / F$_a$ and,
since  the FBZ area varies as F$_a$ under pressure, the relevant
lattice parameter is proportional to 1 / (F$_a$)$^{1/2}$, yield:

\begin{equation}
\label{Eq_A} A \propto \left(\frac{m^*_a(P)}{m^*_a(P =
0)}\right)^2\left(\frac{F_a(P)}{F^*_a(P = 0)}\right)^{5/2}
\end{equation}

As can be observed in Fig. \ref{fig_mc_TD_B0(P)}a, Eq. \ref{Eq_A}
accounts for the data within the error bars. Such a behaviour
could suggest a reduction of electron correlations under pressure,
in line with a Brinkman-Rice scenario \cite{Me00}. However, it
should be mentioned that the product A$\times$T$_0^2$ (where T$_0$
is the coherence temperature, above which the T$^2$ law is no more
valid) should be roughly pressure-independent within this picture
which is not the case. Indeed, according to the data of Fig.
\ref{fig_RT}, it decreases by about a factor of two from ambient
pressure to 1.1 GPa which suggests that electron correlations may
not be the only contribution to the observed behaviour. A similar
conclusion is derived from the data of
$\kappa$-(ET)$_2$Cu[N(CN)$_2$]Br \cite{St05}.

Besides the effective mass, the two main ingredients entering the
field dependence of the oscillation amplitude are the MB field and
the Dingle temperature. Since the $a$ orbits only involve Bragg
reflections, both T$_D$ and B$_0$\footnote{Two different MB gaps
are observed in Fig. \ref{fig_SF}. As discussed in Ref.
\cite{Vi03}, only a value very close to their arithmetic mean, can
be derived from the field-dependent data.} can be derived from
this component \cite{Vi03,Au05}. The MB field value, consistently
deduced from both the $a$ and $b$ components within the error
bars, strongly decreases as the applied pressure increases as
reported in Fig. \ref{fig_mc_TD_B0(P)}c. This behaviour is
consistent with the data relevant to the MB orbit $\beta$ reported
for few charge transfer salts \cite{Ca94,Br95}. T$_D$ has been
derived from the $a$ and $2a + \delta$ components, adopting the
B$_0$ value derived from $a$ in the latter case (see Fig.
\ref{fig_mc_TD_B0(P)}b). As already reported \cite{Vi03}, a
slightly negative value is obtained at ambient pressure for  $2a +
\delta$. This can be due to a contribution of the frequency mixing
phenomenon to this component. A lack of accuracy in the
determination of B$_0$ (large error bars are observed in Fig.
\ref{fig_mc_TD_B0(P)}c) can also contribute to this result since
the $2a + \delta$ orbits involves four tunnellings (t$_{2a +
\delta}$ = 4 in Eq. \ref{RMB}) which makes the deduced T$_D$ value
very sensitive to B$_0$. A sizeable increase of the scattering
rate, deduced from the field dependence of both the \emph{a} and
2\emph{a}+$\delta$ components, is observed under pressure. Indeed,
T$_D$ which is close to 0 at ambient pressure increases up to
about 4 K at 1.1 GPa which corresponds to $\tau_D$ = 0.3 ps.
Pressure-induced defects seems the most plausible explanation for
this behaviour. However, it must be recalled that this high
$\tau_D$ value decreases back as the pressure is released which
rules out any sample degradation due to the pressurization
process. The increase of the scattering rate is in line with the
pressure-induced resistance maximum observed in Fig. \ref{fig_RT}.
However, as above discussed, the decrease of its amplitude above
0.5 GPa cannot be interpreted on this basis. An interplay between
the pressure-induced decrease strength of the electron
correlations and the increase of scattering rate, the latter being
in this case assumed to be mainly controlled by electron-phonon
interaction, could therefore be considered. Within this picture,
it should also be assumed that the variations of the interlayer
scattering rate, involved in the interlayer resistance, reflect
that of the in-plane scattering rate which is probed by SdH
oscillations. In any case, the observed pressure dependence of the
Dingle temperature is at variance with the data of most ET-based
salts for which T$_D$ remains unchanged under pressure
\cite{Br95,Ha03,Vi08}.

Finally, let us examine the pressure dependence of the frequency
mixing that correspond to so called "forbidden orbits". They could
arise from both chemical potential oscillation \cite{mu} and
MB-induced Landau level broadening \cite{LLB}. As mentioned in
Ref. \cite{Ki02}, applied pressure leads to an increase of the
transverse interactions and therefore to a damping of the chemical
potential oscillation. Oppositely, as can be observed in Fig.
\ref{fig_TF}, the amplitude of some components such as $\Delta +
a$ or $b + a$ remains large, or even increases, as the applied
pressure increases. This behaviour suggests that the
field-dependent chemical potential oscillation is not the dominant
contribution to the development of the observed frequency
combinations. This point is in line with recent calculations
\cite{Fo08} which indicate that, contrary to the case of numerous
ET salts which are uncompensated metals, chemical potential
oscillation in compensated orbits systems is strongly damped. As a
consequence, Landau level broadening is likely the main source of
the observed frequency combinations in compensated orbits
networks.

\section{Summary and conclusion}

The SdH spectra of the charge transfer salt
(ET)$_8$[Hg$_4$Cl$_{12}$(C$_6$H$_5$Br)]$_2$ exhibits many Fourier
components corresponding to linear combinations of three basic
frequencies (F$_a$, F$_\delta$ and F$_\Delta$), in agreement with
band structure calculations (see Fig. \ref{fig_SF}). Most of these
frequency combinations, which correspond to "forbidden orbits" in
the framework of the semiclassical model, keep a large amplitude
in all the studied pressure range. This feature likely rules out
chemical potential oscillation as a dominant contribution to their
origin.

The pressure sensitivity of the FBZ area is similar to that of
most of the ET salts. The area of the $\Delta$ piece and of the
compensated orbits (\emph{a}) follow the same pressure dependence
as that of the FBZ area while the area of the $\delta$ piece
decreases under pressure. The measured electronic properties are
significantly modified as the applied pressure increases. Either
additional frequencies or shifts in few of the frequency
combinations are observed in the high pressure range. Therefore,
some change of the FS topology as the pressure increases cannot be
excluded. Higher pressures are needed to check this hypothesis.

The zero-field interlayer resistance follows a T$^2$ behaviour at
low temperature with a pressure-dependent prefactor A, in
agreement with the predictions for a strongly correlated Fermi
liquid \cite{Mo95}. Since it strongly decreases under pressure, it
could be inferred that the effect of electron correlations, which
are important at ambient pressure decreases, under pressure. This
latter point is in agreement with data relevant to e.g.
$\kappa$-(ET)$_2$Cu[N(CN)$_2$]Cl \cite{Li03}. However, within this
picture, the amplitude of the resistance maximum should also
continuously decrease as the applied pressure increases, which is
not the case. Oppositely, it increases in the pressure range below
about 0.5 GPa. In line with SdH data reported for
$\kappa$-(ET)$_2$Cu[N(CN)$_2$]Br \cite{St99}, the increase of the
amplitude of the resistance maximum in the low pressure range
could be attributed to the sizeable increase of the scattering
rate, deduced from the SdH data, as the applied pressure
increases. Within this picture, the pressure dependence of the
interlayer resistance would mainly result from the interplay
between the pressure sensitivity of the scattering rate and of the
strength of electron-correlations.

\begin{acknowledgement}
 This work was supported by Euromagnet under the European Union contract
R113-CT-2004-506239, MEC-Spain (Project FIS2006-12117-C04-01,
CSD2007-00041), Generalitat de Catalunya (Project 2005 SGR 683),
the French-Russian exchange program between CNRS and RAS (number
21451) and by RFBR grants 09-02-00899 and 08-03-00480. We thank
Jean-Yves Fortin, Sergei I. Pesotskii and Geert Rikken for
interesting discussions.
 \end{acknowledgement}

% BibTeX users please use
%\bibliographystyle{unsrt}
%\bibliographystyle{apsrev}
%\bibliography{biblio}

\end{document}